\documentclass[aps,nofootinbib]{revtex4}
\usepackage{amssymb}
\usepackage{amsmath}
\usepackage{graphicx}
\usepackage{eurosym}
\usepackage{dsfont}
\usepackage{amsmath}
\usepackage{amssymb}
\usepackage{amsthm}
\usepackage{stmaryrd}
\usepackage{mathrsfs}
\usepackage{textcomp}
\usepackage{amsfonts}
\usepackage{txfonts}
\usepackage{color}
\def\bel{\begin{equation}\label}
\def\ee{\end{equation}}
\def\beq{\begin{eqnarray}}
\def\eeq{\end{eqnarray}}

\begin{document}

\title{{\LARGE \textbf{Unified description for $\kappa-$deformations of orthogonal groups}}}
\author{A. Borowiec  \footnote{andrzej.borowiec@ift.uni.wroc.pl}}
\affiliation{Institute for Theoretical Physics,pl. M. Borna 9, 50-204 Wroc{\l}aw, Poland}

\author{A. Pacho{\l} \footnote{pachol@hi.is}}
\affiliation{Science Institute, University of Iceland, Dunhaga 3, 107 Reykjavik, Iceland}

\begin{abstract}
In this paper we provide universal formulas describing Drinfeld-type quantization of inhomogeneous orthogonal groups determined by a metric tensor of an arbitrary signature living in a spacetime
of arbitrary dimension. The metric tensor does not need to be in diagonal
form and $\kappa-$deformed coproducts are presented in terms of classical generators. It opens the possibility for future applications in deformed general relativity. The formulas depend on the choice of an additional vector field which parametrizes classical $r-$matrices. Non-equivalent deformations are then labeled by the corresponding type of stability subgroups. For the
Lorentzian signature it covers three (non-equivalent) Hopf-algebraic
deformations: time-like, space-like (aka tachyonic) and light-like
(a.k.a. light-cone) quantizations of the Poincar\'{e} algebra. 
Finally the existence of the so-called Majid-Ruegg (non-classical) basis is
reconsidered.
\end{abstract}

\maketitle

\section{Introduction}

Deformations of relativistic symmetries have been fruitful for the
description of quantum symmetries governing physics at the Planck scale.
Such quantum deformations of spacetime symmetries are described within the
Hopf algebra language and are controlled by classical $r$ -matrices
satisfying the classical Yang-Baxter (YB) equation: modified or unmodified one. One of the most interesting deformations, from the point of view of physical
applications, the so-called $\kappa $-deformation has been found in \cite{Luk1,LNR,Luk2,MR}. The deformation  parameter corresponds to the Planck Mass; its inverse defining fundamental length can be considered
as a quantum gravity scale. The $r$ -matrix for the $\kappa $-deformation of Poincar\'{e}
algebra is given then by $r=M_{0i}\wedge P^{i}$  and it satisfies the
modified  (inhomogeneous) Yang-Baxter equation (MYBE): $[[r,r]]=M_{\mu \nu }\wedge P^{\mu }\wedge P^{\nu }$. The $\kappa $%
-Poincar\'{e} Hopf algebra  constitutes the deformed
symmetry of the $\kappa $-Minkowski algebra \cite{MR,Z} which is a quantum version of the standard Minkowski spacetime. The
$\kappa $-Minkowski spacetime has been mostly studied in the so-called
time-like version of $\kappa$-deformation, distinguishing the 'time' coordinate as the
quantized one. The r-matrix mentioned above corresponds to this case
\footnote{The corresponding classification of quantum deformations (complete for Lorentz and almost-complete for Poincar\'{e} algebras) has been performed in Ref. \cite{Zakrzewski} 
(see also dual matrix quantum group version in \cite{PodlesWoronowicz}).}.
Another option is the so-called light-like (null-plane) deformation corresponding to null-vectors,
which was firstly considered in \cite{BallesterosPLB351} (then also in \cite%
{BallesterosPLB391}, \cite{LukMinnMozrz}) with quantum-deformed direction on
the light cone $\left( x^{+}=x^{0}+x^{3}\ \right) $ and with the corresponding
symmetry the so-called 'null-plane quantum Poincar\'{e} Lie algebra'. It was
inspired by the central problem of quantum relativistic systems in the
Hamiltonian formulation, which has been studied for the null-plane
evolution. In this case the information provided by the Poincar\'{e}
invariance splits into a dynamical and kinematical part which is also the
case after the deformation. One of the advantages of the deformation of this
type is that it is triangular i.e. it can be described by the classical r-matrix satisfying classical
Yang-Baxter equation (CYBE) and the twisting element satisfying two-cocycle condition do exist \cite{Mudrov}.
Moreover, the differential calculus for the null-plane $\kappa -$Minkowski is
shown to be bicovariant and four-dimensional \cite{0307038}, which has been
proved to be impossible to built for other kinds of $\kappa $-deformations
(i.e. time- and space-like) \cite{Sitarz}. It was also shown \cite{Przanowski} that after suitable (nonlinear) change of basis the quantum
algebra presented in \cite{BallesterosPLB351} can be identified with the $%
\kappa $-deformation, given in \cite{koma95} for the choice of $g_{00}=0$.

Till now the most popular form of presentation of quantum $\kappa $-Poincar\'{e} algebra is
the one which uses formulas for deformed coproducts found for the first time in \cite{MR} (with the primitive energy generator $P_0$). The corresponding system of generators, known also as Majid-Ruegg or bicrossproduct basis, satisfy classical commutation relations between Lorentzian generators and deformed ones in the boosts-momenta sector.
In contrast our formulae for quantized coproducts are written entirely in the classical Lie algebra basis \footnote{It has been demonstrated in \cite{AB_AP_bicross} that the classical basis is related with bi-crossed product construction as well.}. Some formulas for coproducts can be found in different (realization-dependent) context in \cite{GNbazy}, \cite{Group21}, \cite{Wess_1}, \cite{1110.0944}, see also \cite{Meljanac0702215}. The $\kappa $-Poincar\'{e} algebra combined with the non-orthogonal form of the metric tensor was originally studied in \cite{koma95} in Majid-Ruegg basis and later with extended analysis, e.g. in \cite{0307038}. A passage from the Majid-Ruegg into the classical basis, which provides the so-called Drinfeld quantization map, has been a subject of investigations in various context \cite{KosLuk}, \cite{Kos}. Particularly, the explicit formulas expressing classical basis in terms of bicrossproduct one have been obtained therein.  Similarly, the null-plane deformation has been originally obtained and investigated in the basis inherited from the so-called deformation embedding method \cite{BallesterosPLB351}. The classical Lie algebra basis in this context has not been explored yet.%

Our aim in this paper is to provide a unified description for  $\kappa $-deformed coproducts of classical Poincar\'{e} generators  characterizing various $\kappa$-deformations according to the Zakrzewski classification scheme \cite{Zakrzewski}.
The formulas depend on the choice of an external vector field $\tau$ which
parameterizes classical r-matrices. Non-equivalent deformations are then labeled (classified) by the
corresponding stability groups of $\tau$. 
The metric tensor can take the form of arbitrary symmetric and non-degenerate matrix.
For example for the Lorentzian signature (in arbitrary dimension) one can distinguish three different
quantizations: time-like, space-like and light-like. The corresponding orbits are characterized by the following stability
subgroups in $S\!O(D-1, 1)$: $S\!O(D-1), IS\!O(D-2), S\!O(D-2,1)$ respectively. This form of the unified description is particularly important from the point of view of future applications in deformations of general relativity \cite{Wess} where the
metric might be a function of the coordinates \cite{Majid-Beggs} and/or in the
so-called relative locality where it might live on the momentum space \cite{GAC}.

The universal formulas are followed by the example of orthogonal $D=1+(D-1)$ decomposition, which is suitable for
non-null $\tau$.  This case allows for the change of system of generators into the well-known Majid-Ruegg (nonlinear) basis. Another example is the null-vector case ($\tau^2=0$) which provokes orthogonal $2+(D-2)$ decomposition.
This (a.k.a. null-plane) case admits additionally Drinfeld twist, due to the fact that r-matrix satisfies CYBE.
The universal formulas for coproducts coincide (up to quantum $R-$matrix) with the twisted ones.
Also the partial analog of Majid-Ruegg basis can be found in that case. We
finish this paper with some conclusions and perspectives.

\section{Preliminaries and notation}

Let $V$ be a vector space (spacetime) of arbitrary dimension $D$ equipped
with the metric tensor $g$ of arbitrary signature. Let denote the (special) orthogonal group of $g$
as $S\!O\left( g\right) =\{\Lambda\in GL(V):
\Lambda^T g \Lambda=g, \det \Lambda=1\}$ and the corresponding inhomogeneous orthogonal group as $IS\!O\left(
g\right) $ - Poincar\'{e} group \footnote{In fact, we should restrict ourselves to the connected component of unity instead of the full $SO(g)$. For example the special Lorentz group $SO(1, 3)=SO^\uparrow(1, 3)\cup SO^\downarrow(1, 3)$ has two connected components.}.
Adopting typical relativistic notation one chooses the basis $\{e_{\mu}\}_{\mu =0}^{D-1}$ and introduces the components of the metric: $g_{\mu
\nu}=g\left( e_{\mu },e_{\nu }\right) $.
There always exists an orthonormal basis
$\{e _{a}\}$ and a vierbein matrix $\xi_{a}^{\mu } $ which diagonalizes the metric and then $\ g_{\mu \nu }=\xi _{\mu
}^{a}\xi _{\nu }^{b}\eta _{ab}$ with the diagonal elements $\eta_{aa}=\pm 1$. Therefore we are used to write $IS\!O(p, q)$
in order to distinguish between positive and negative diagonal entries: $p+q=D$.

The Lie algebra $iso\left( g\right) $ as an infinitesimal form of this group admits the Lie algebra basis
$\{M_{\mu \nu}$, $P_{\mu }\}$ adopted to the basis $\{e_{\mu}\}_{\mu =0}^{D-1}$ in  $V$. It consists of the familiar  commutation relations:
\begin{eqnarray}
\lbrack M_{\mu \nu },M_{\rho \lambda }] &=& i(g_{\mu \lambda }M_{\nu \rho
}-g_{\nu \lambda }M_{\mu \rho }+g_{\nu \rho }M_{\mu \lambda }-g_{\mu \rho
}M_{\nu \lambda }),  \label{MM} \\
\lbrack M_{\mu \nu },P_{\rho }] &=& i(g_{\nu \rho }P_{\mu }-g_{\mu \rho
}P_{\nu })\quad, \quad \lbrack P_{\mu },P_{\lambda }] = 0.  \label{PP}
\end{eqnarray}%
The universal enveloping algebra $U_{iso\left( g\right) }$ of this Lie
algebra can be equipped with a primitive Hopf algebra structure, which can later be quantized within the Drinfeld formalism. Because of this, one requires the extension to the
formal power series $U_{iso\left( g\right) }[[\frac{1}{\kappa}]]$ (see e.g. \cite{Drinfeld,ChP} for more details).
This associative and unital algebra has a
quadratic Casimir element (aka Casimir of mass) defined as
$C_{g}=P_{\mu }P^{\mu }=g^{\mu \nu }P_{\mu }P_{\nu }=C \in
U_{io\left( g\right) }$ (a central  element of $U_{iso\left( g\right) }$)
which plays very important role in physics, it represents the mass. It takes
a constant numerical value in any irreducible representation.

However, from the point of view of physical applications e.g. at the Planck scale we are
interested in quantum deformations. As is well known quantum groups are
quantizations of Poisson Lie groups determined by Lie bialgebra structures on the
corresponding Lie algebras. They are described by classical $r-$matrices satisfying Yang-Baxter equation.
In the case of orthogonal groups an interesting class of  r-matrices has been found in \cite{Zakrzewski}
(see also \cite{Stachura}).   For any (non-zero) vector
$\tau =\tau ^{\mu }e_\mu\in V$ one defines the corresponding $r-$matrix
\begin{equation}\label{z1}
r_{\tau } ={\tau }^{\alpha }M_{\alpha \mu }\wedge P^{\mu }\equiv
\tau\llcorner\, \Omega\ ,
\end{equation}
where $\Omega =M_{\mu \nu }\wedge P^{\mu }\wedge P^{\nu }$ is known to be the only invariant
element in $\wedge^3iso(g)$ and $\tau\llcorner$ is used for contraction with the vector $\tau$.
It appears that the Schouten bracket gives
\begin{equation}\label{z2}
\left[ \left[ r_{\tau },r_{\tau }\right] \right] =-g(\tau,\tau)\, \Omega.
\end{equation}
This implies two possibilities:

I. $\tau^2\equiv g(\tau, \tau)\equiv \tau^\mu\tau_\mu\neq 0$ for which the corresponding
r-matrix satisfies MYBE (Modified Yang-Baxter Equation).
It will provide the so-called standard (a.k.a. Drinfeld-Jimbo) quantization with the quasi-triangular quantum $R-$matrix.

II. $\tau^2=0$ (provided non-Euclidean signature) with $r_\tau$ satisfying CYBE (Classical
Yang-Baxter equation). In this case one deals with the non-standard (a.k.a. twisted) triangular deformation.

We consider a stability group  $G_\tau$ of the vector $\tau$, as a subgroup  which leaves
the vector $\tau$ invariant under the natural action of $SO(g)$ in $V$.
Isomorphism classes of stability groups classify the type of orbits.
According to the general formalism developed in \cite{Zakrzewski} they can also be
used to single out the non-equivalent deformations labeled by $\tau$.
Regarding the possible orbit types for the non-trivial vector $\tau\neq 0$ in $(V, g)$, assuming generic $(p,q)$ signature, one can encounter two main situations:

A) $\tau^2\neq 0$ and there is a basis $\{e_0=\tau, e_i\}_{i=1}^{D-1}$ in $V$ such that $g_{0i}=0$. This basis provides the so-called $1+(D-1)$ orthogonal decomposition. The stability subgroup is a homogeneous orthogonal group $S\!O(g_{ij})$ in
$D-1$ dimensions. The signature of the metric $g_{ij}$ indicates the orbit type.

B) $\tau^2=0$ (provided we have a non-Euclidean signature). There is a basis $\{e_0=\tau, e_{D-1}=\tilde\tau, e_a\}_{a=1}^{D-2}$ in $V$ such that $g_{00}=g_{D-1\,D-1}=g_{0a}=g_{D-1\,a}=0$ and $g_{0\,D-1}=g_{D-1\,0}=1$. This basis is called a light-cone basis and it provides the so-called $2+(D-2)$ orthogonal splitting. The two-dimensional Lorentzian space with anti-diagonal metric is spanned by two light-like vectors $\tau,\tilde{\tau}$. Again the signature of the metric $g_{ab}$ indicates the orbit type.
The stability subgroup is an inhomogeneous orthogonal group $ISO(g_{ab})$ in $D-2$ dimensions in this case. In other words if $1\leq p,q\leq D-1$ one distinguishes three cases:  either $S\!O(p-1,q)$ or $S\!O(p,q-1)$ or  $IS\!O(p-1,q-1)$. Particularly, for the Lorentzian signature  one recovers the well-known cases:

i) $\tau $ is a time-like vector, with Euclidean $G_{\tau }\cong S\!O(D-1)$  as a stability
group. It corresponds to the original $\kappa$-deformation with $r_\tau=M_{0i}\wedge P^i$;

ii) $\tau ^{2}=0$, i.e. $\tau$ is a null-vector (light-cone deformation). In this case the stability group $%
G_\tau \cong E(D-2)\equiv IS\!O(D-2)$ is an inhomogenous Euclidean group;

iii) $\tau$ is a space-like vector (tachyonic deformation) with $G_\tau\cong S\!O(D-2,1)$ being $D-1$ dimensional Lorentz group.

\section{Unified description for $\protect\kappa-$ deformations}

For a given pair $(g_{\mu \nu },\tau ^{\lambda })$ the deformed coproducts
implemented by the vector $\tau $ (in fact, by the corresponding classical $%
r-$matrix $r_{\tau }$), when written in classical
generators satisfying the commutation relations (\ref{MM})--(\ref{PP})~\footnote{%
For realization-dependent version see e.g. \cite{Wess_1,1110.0944,Meljanac0702215}.}, take the following form:
\begin{eqnarray}
\Delta _{\tau }\left( P_{\mu }\right) &=&P_{\mu }\otimes \Pi _{\tau
}+1\otimes P_{\mu }-\frac{\tau _{\mu }}{\kappa }P^{\alpha }\Pi _{\tau
}^{-1}\otimes P_{\alpha }-\frac{\tau _{\mu }}{2\kappa ^{2}}C_{\tau }\Pi
_{\tau }^{-1}\otimes P_{\tau }  \label{cpP} \\
\Delta _{\tau }\left( M_{\mu \nu }\right) &=&M_{\mu \nu }\otimes 1+1\otimes
M_{\mu \nu }+\frac{1}{\kappa }P^{\alpha }\Pi _{\tau }^{-1}\otimes \left(
\tau _{\nu }M_{\alpha \mu }-\tau _{\mu }M_{\alpha \nu }\right)  -\frac{1}{2\kappa ^{2}}C_{\tau }\Pi _{\tau }^{-1}\otimes \left( \tau _{\mu
}M_{\tau \nu }-\tau _{\nu }M_{\tau \mu }\right)  \label{cpM}
\end{eqnarray}%
where $\tau _{\mu }$ denote covariant components of $\tau $ with respect to
the metric $g_{\mu \nu }$ and $P_{\tau }=\tau ^{\mu }P_{\mu },M_{\tau
\lambda }=\tau ^{\alpha }M_{\alpha \lambda }$. In order to preserve a
compact form for the formulas (\ref{cpP})--(\ref{cpM}) we have also
introduced the following notation (extending our previous notation from \cite%
{AB_AP_JPA}):
\begin{eqnarray}
\Pi _{\tau } &=&\frac{1}{\kappa }P_{\tau }+\sqrt{1+\frac{1}{\kappa ^{2}}\tau
^{2}C}\quad ,\quad \Pi _{\tau }^{-1}=\frac{\sqrt{1+\frac{1}{\kappa ^{2}}\tau
^{2}C}-\frac{1}{\kappa }P_{\tau }}{1+\frac{1}{\kappa ^{2}}\left( \tau
^{2}C-P_{\tau }^{2}\right) }  \label{d1} \\
\tau ^{2}C_{\tau } &=& \kappa ^{2}\left( \Pi _{\tau }+\Pi _{\tau }^{-1}-2+%
\frac{1}{\kappa ^{2}}\left( \tau ^{2}C-P_{\tau }^{2}\right) \Pi _{\tau
}^{-1}\right)  \label{d2}
\end{eqnarray}%
The left (and right) hand side of the last equation vanishes when $\tau
^{2}=0$.  Further calculations give rise to
\begin{equation}
P_{\tau }=\frac{\kappa }{2}\left( \Pi _{\tau }-\Pi _{\tau }^{-1}\left( 1+%
\frac{1}{\kappa ^{2}}\left( \tau ^{2}C-P_{\tau }^{2}\right) \right) \right)
\quad ,\quad \tau ^{2}C_{\tau }=2\kappa ^{2}\left( \sqrt{1+\frac{1}{\kappa
^{2}}\tau ^{2}C}-1\right)  \label{d3}
\end{equation}
We would like to point out that all formulas considered so far are valid also in the
null-vector case, i.e. as $\tau ^{2}=0$. (For $\tau\equiv 0$ one recovers the primitive
undeformed coproducts).
We shall specify later on the expression for $C_{\tau }$ for this
particular case. At the moment one can observe that the inverse to (\ref{d3}),
formula
\begin{equation}\label{inv}
    C=C_\tau\left(1+\frac{\tau^2}{4\kappa^2}C_\tau \right)
\end{equation}
strongly suggests $C_\tau=C$ for $\tau^2=0$.

Following the Drinfeld formalism of quantum groups all equalities presented here
are understood in the sense of formal power series in one (undetermined)
variable ${\frac{1}{\kappa }}$, for example \footnote{For the standard (i.e. Drinfeld-Jimbo-type) deformation one can always switch to the so-called q-analogue version with all infinite series hidden in the one additional generator. In the case of $\kappa-$Poincare it is $\Pi _{\tau }^{-1}$   which solves a specialization problem for $\kappa$, for details see \cite{AB_AP_JPA}.}
\begin{equation}
\sqrt{1+\frac{1}{\kappa ^{2}}\tau ^{2}C}=\sum_{n\geq 0}\frac{(\tau ^{2})^{n}%
}{\kappa ^{2n}}\,\binom{1/2}{n}\,C^{n} = 1+\frac{\tau^2C}{2\kappa^2}+\sum_{n\geq 2}\frac{(-1)^{n-1}(\tau ^{2})^{n}%
(2n-3)!!}{2^nn!\kappa ^{2n}}C^n\ , \label{sqrt}
\end{equation}%
where $\binom{1/2}{n}=\frac{1/2(1/2-1)\ldots (1/2-n+1)}{n!}$ are binomial
coefficients. $\mathcal{C}_{\tau }$ is a central element in $U_{iso(p,q)}[[{%
\frac{1}{\kappa }}]]$ which in the classical limit $\kappa \mapsto \infty $
gives $C$ and, therefore, plays the role of deformed Casimir operator. From the above
ones one calculates
\begin{equation}
\Delta _{\tau }(\Pi _{\tau })=\Pi _{\tau }\otimes \Pi _{\tau }\quad ,\quad \Delta
_{\tau }(\Pi _{\tau }^{-1})=\Pi _{\tau }^{-1}\otimes \Pi _{\tau
}^{-1},  \label{Pi3}
\end{equation}%
as well as ($\tau ^{2}\neq 0$)
\begin{equation}
\ \Delta _{\tau }\left( \sqrt{1+\frac{1}{\kappa ^{2}}\tau ^{2}C}\right) =%
\sqrt{1+\frac{1}{\kappa ^{2}}\tau ^{2}C}\otimes \Pi _{\tau }-\frac{1}{\kappa
}\Pi _{\tau }^{-1}\otimes P_{\tau }+\frac{\tau ^{2}}{\kappa ^{2}}P^{\alpha
}\Pi _{\tau }^{-1}\otimes P_{\alpha }-\frac{\tau ^{2}}{\kappa }P_{\tau }\Pi
_{\tau }^{-1}\otimes P_{\tau }
\end{equation}%
Finally, in order to complete the Hopf algebra structure we set the
classical counit ($\epsilon (1)=1,\epsilon (P_{\lambda })=\epsilon (M_{\mu
\nu })=0$) and deformed antipodes:
\begin{eqnarray}
S_{\tau }\left( P_{\mu }\right) &=&-\left( P_{\mu }+\frac{\tau _{\mu }}{%
\kappa }\left( C+\frac{1}{2\kappa }P_{\tau }\,C_{\tau }\right) \right) \Pi
_{\tau }^{-1}\quad,\qquad
S_{\kappa }(\Pi _{\tau })=\Pi _{\tau }^{-1} \label{SP}\\
S_{\tau }\left( M_{\mu \nu }\right) &=&-M_{\mu \nu }+\frac{1}{\kappa }%
P^{\alpha }\left( \tau _{\nu }M_{\alpha \mu }-\tau _{\mu }M_{\alpha \nu
}\right) +\frac{1}{2\kappa ^{2}}C_{\tau }\left( \tau _{\nu }M_{\tau \mu
}-\tau _{\mu }M_{\tau \nu }\right)\label{SM}
\end{eqnarray}
Moreover, the square of the antipode ( \ref{SP} ), ( \ref{SM} ) is given by a similarity transformation $S^2(X)=\Pi _{\tau }^{D-1}X\Pi _{\tau }^{1-D}$ (cf. \cite{1110.0944}). Such deformed Hopf algebraic structure will be denoted as $%
U_{iso(g)}^{\,\tau }[[{\frac{1}{\kappa }}]]$.\\
\textit{Remark 1}:\\
It is worth to underline that these universal formulas describe $\kappa$-Poincare Hopf algebra not only  in different Lie algebra basis induced by different  basis in the underlying vector space $V$ but also provide the different types of deformations. This can be seen from the well-known formula
\begin{equation}
\lim_{\kappa\shortrightarrow\infty} \kappa(\Delta_\tau-\Delta^{op}_\tau)(X)=[\Delta_0(X)\,,r_\tau]
\end{equation}
relating deformed coproducts with the corresponding classical $r-$matrices. Here $\Delta_0(X)=X\otimes 1+1\otimes X$ denotes primitive (undeformed) coproduct for $X\in iso(g)$ and $\Delta^{op}$ stands for the opposite coproduct with flipped legs.
The right hand side of the last equation defines cobracket determining Lie bialgebra structures on $iso(g)$. Therefore our coproducts can be considered as their quantization. The following further comments are now in order.\newline
\textit{Remark 2:}\newline
One should notice that the expression $\tau ^{2}\,C$ is independent of the
sign convention for $g$: the change $g\rightarrow -g$ gives rise to $\tau ^{2}\rightarrow
-\tau ^{2}$ and $C\rightarrow -C$.\newline
\textit{Remark 3:}\newline
Re-scaling at the same time $\tau \rightarrow s \tau $ and $\kappa
\rightarrow s \kappa $ for any real parameter $s $ leaves
coproducts (\ref{cpP})-(\ref{cpM}) invariant. Notice that neither $\tau $ nor $%
\kappa $ are present in the commutation relations (\ref{MM})--(\ref{PP}).
For this reason (except the case $\tau ^{2}=0$) one can assume that the
vector $\tau $ is normalized, i.e. $\tau ^{2}=\pm 1$ provided $\tau ^{2}\neq
0$.\newline
\textit{Remark 4:}\newline
 Consider the well-known $\kappa $-Minkowski (quantum) algebra  $\mathcal{M}_{\tau }$ as a unital associative algebra generated by the noncommutative spacetime coordinate generators $\hat x^\mu$ modulo the following relations \cite{koma95}
\begin{equation}\label{kM}
\left[ \hat x^{\mu }, \hat x^{\nu }\right] =\frac{i}{\kappa }\left( \tau ^{\mu }\hat x^{\nu
}-\tau ^{\nu } \hat x^{\mu }\right)
\end{equation}
where $\tau ^{\mu }$ is a fixed four-vector from $V$; $\mu,\nu =0, 1,\ldots, D-1$. This algebra becomes a Hopf module algebra (see e.g. \cite{AB_AP_sigma} for necessary definitions) with respect to the $\kappa-$deformed Hopf algebra structure (\ref{cpP})-(\ref{cpM}). It means that the relation (\ref{kM}) is preserved under the module action $\triangleright$, provided classical action of the classical Poincar\'{e} generators (\ref{MM})-(\ref{PP}) on the $\kappa -$%
Minkowski coordinates (\ref{kM}):
\begin{equation}\label{kM1}
P_{\mu }\triangleright \hat x^{\nu }=-\imath \delta _{\mu }^{\nu }\qquad,
\qquad M_{\mu \nu }\triangleright \hat x^{\rho }=-i\left( \hat x_{\mu }\delta _{\nu
}^{\rho }-\hat x_{\nu }\delta _{\mu }^{\rho }\right)
\end{equation}%
To this aim one requires the compatibility condition (a.k.a. generalized Leibniz rule):
\begin{equation}\label{kM2}
L\triangleright \left( \hat x^{\mu }\cdot \hat x^{\nu }\right) =\left( L_{\left(
1\right) }\triangleright \hat x^{\mu }\right) \cdot \left( L_{\left( 2\right)
}\triangleright \hat x^{\nu }\right)
\end{equation}%
where, for simplicity, we have used Sweedler-type notation for the coproduct: $\Delta_\tau(L)=L_{\left( 1\right)}\otimes L_{\left( 2\right)} $ for $L\in \{M_{\mu\nu}, P_\rho\}$. Using a smash product construction one can unify spacetime and symmetry generators (see e.g. \cite{AB_AP_sigma}) into one algebra with quantum Hopf algebroid structure \cite{BKJMP}.\\
\textit{Remark 5:}\newline
It is well known that the real algebras $U_{iso(p,q)}$ of different signatures $(p, q)$ can be viewed  as different real forms determined on the same complex algebra $U_{iso(D,\mathds{C})}$. These  real forms are represented by the corresponding $*$-conjugation. The standard and convenient way to establish appropriate conjugations is by the choice of a Lie algebra basis composed of self-conjugate (Hermitean) elements. One can observe that the basis (\ref{MM})-(\ref{PP}) is compatible with Hermitean conjugation and it can be used for determining the corresponding real forms. Thus the universal coproducts (\ref{cpM}) are compatible with the signature implemented $*-$ conjugations in the following sense: $$\Delta_\tau(a)^*=\Delta_\tau(a^*); \quad (a\otimes b)^*=a^*\otimes b^*; \quad a, b\in U_{iso(D,\mathds{C})}$$
 Similarly, the relation (\ref{kM}) can be considered as providing the real structure on the complex module algebra  $\mathcal{M}_{\tau }$ provided that the vector $\tau$ remains to be real.\\
\textit{Examples:}\newline
Take the diagonal metric $\eta _{\mu \nu }=\eta ^{\mu \nu }=(-,+,+,+)$ of the
Lorentzian signature in $D=4$ dimensions. Three different choices: $%
_{1}\!\tau ^{\mu }=(1,0,0,0)$, $_{1}\!\tau ^{2}=-1$;\qquad\ $_{2}\!\tau
^{\mu }=(0,0,0,1)$, $_{2}\!\tau ^{2}=1$ \qquad and $_{3}\!\tau ^{\mu
}=(1,0,0,1)$, $_{3}\!\tau ^{2}=0$ provide three different (non-equivalent)
Hopf algebraic structures on $U_{iso(1,3)}[[{\frac{1}{\kappa }}]]$: the original $\kappa $, tachyonic and light-cone deformations, respectively
(see also \cite{LukMinnMozrz},\cite{Przanowski},\cite{koma95},\cite{hep-th0203182} for earlier works in this context). We shall denote them $%
U_{iso(1,3)}^{SO(3)}[[{\frac{1}{\kappa }}]]$, $\quad U_{iso(1,3)}^{\,E(2)}[[{%
\frac{1}{\kappa }}]]$ \quad and $U_{iso(1,3)}^{\,SO(1,2)}[[{\frac{1}{\kappa }%
}]]$ correspondingly. These examples will be treated in more detail in the next subsections.

Yet another example can be considered by taking the diagonal metric $\eta_{\mu\nu}=\eta^{\mu\nu}=(+,-,+,-)$ of neutral (Kleinian) signature in $D=4$ dimensions. The choice $\tau^\mu = (1, 1, 1, 1)$ ($\tau^2=0$) provides a new type of deformation of $U^{}_{iso(2,2)}$. Equivalently one can take more convenient basis with the metric \footnote{This indicates Lie algebra isomorphism $iso(2,2)\cong iso(1,1)\oplus iso(1,1)$.}
$$g_{\mu \nu }=g^{\mu \nu }=\left(
\begin{array}{cccc}
0 & 1 & 0 & 0 \\
1 & 0 & 0 & 0 \\
0 & 0 & 0 & 1 \\
0 & 0 & 1 & 0%
\end{array}%
\right) $$
being a direct product of the two light-cone metrics, and we have $\tau^\mu = (1, 0, 1, 0)$. In this basis one sees Hopf algebra isomorphism $U^{\,\tau}_{iso(2,2)}[[{1\over\kappa}]]\cong U^{\,_1\!\tau}_{iso(1,1)}[[{1\over\kappa}]]\otimes U^{\,_1\!\tau}_{iso(1,1)}[[{1\over\kappa}]]$ with $_1\!\tau=(1, 0)$.

\section{The orthogonal $D=1+(D-1)$ decomposition versus the Majid-Ruegg basis}

Contracting (\ref{cpP})-(\ref{cpM}) with $\tau ^{\mu }$ yields
\begin{eqnarray}
\Delta (P_{\tau }) &=& P_{\tau }\otimes \Pi _{\tau }+\Pi _{\tau }^{-1}\otimes
P_{\tau }-\frac{\tau ^{2}}{\kappa }P^{\alpha }\Pi _{\tau }^{-1}\otimes
P_{\alpha }-\frac{\tau ^{2}}{2\kappa ^{2}}C_{\tau }\Pi _{\tau }^{-1}\otimes
P_{\tau }  \label{aa}\\
\Delta _{\tau }\left( M_{\tau \nu }\right) &=&M_{\tau \nu }\otimes 1+1\otimes
M_{\tau \nu }+\frac{1}{\kappa }P^{\alpha }\Pi _{\tau }^{-1}\otimes \left(
\tau _{\nu }M_{\alpha \tau }-\tau^2 M_{\alpha \nu }\right)-
\frac{\tau^2}{2\kappa ^{2}}C_{\tau }\Pi _{\tau }^{-1}\otimes
M_{\tau \nu }  \label{cpM2}
\end{eqnarray}

Let us study the case of $\tau ^{2}\neq 0$ in more detail (the opposite case
will be the subject of our study in the next section). In fact, without the loss
of generality, one can assume $\tau ^{\mu }=(1,0,\ldots ,0)$. More exactly,
by the choice of the suitable basis $(e_{\mu })_{\mu =0}^{D-1}$ in the vector
space $V$ with $e_{0}=\tau $ and $(e_{i})_{i=1}^{D-1}$ being orthogonal to $%
\tau $ : $g_{00}=\tau ^{2}\,;\quad g_{0i}=g(e_{0},e_{i})=0$. This provides
the orthogonal decomposition $(V,g_{\mu \nu })\cong (\mathds{R}
,g_{00})\times (V^{D-1},g_{ij})$.
Notice that the (D-1) dimensional metric $g_{ij}$ does not need to be in the diagonal form.

In the corresponding Lie algebra basis $\{P_{\tau }, P_i, M_{\tau i}, M_{ij}\}$ the universal coproducts read now as:
\begin{eqnarray}
\quad \Delta _{\tau }\left( P_{\tau }\right)  &=&P_{\tau }\otimes \Pi _{\tau
}+\Pi _{\tau }^{-1}\otimes P_{\tau }-\frac{\tau ^{2}}{\kappa }P^{j}\Pi
_{\tau }^{-1}\otimes P_{j}\label{DPtau} \\
\Delta _{\tau }\left( P_{i}\right)  &=&P_{i}\otimes \Pi _{\tau }+1\otimes
P_{i}\ \quad, \qquad i,j=1,\ldots ,D-1 \\
\Delta _{\tau }\left( M_{ij}\right)  &=&M_{ij}\otimes 1+1\otimes M_{ij} \\
\Delta _{\tau }\left( M_{\tau i}\right)  &=&M_{\tau i}\otimes 1+\Pi _{\tau
}^{-1}\otimes M_{\tau i}+\frac{\tau ^{2}}{\kappa }P^{j}\Pi _{\tau
}^{-1}\otimes M_{ij}\label{DMtau}
\end{eqnarray}%
where $\tau ^{2}$ after normalization can be reduced to $\pm 1$, here we used the
following identity:
\begin{equation*}
\left( 1-\frac{\tau ^{2}}{2\kappa ^{2}}C_{\tau }\Pi _{\tau }^{-1}-\frac{1}{%
\kappa }P_{\tau }\Pi _{\tau }^{-1}\right) =\Pi _{\tau }^{-1}
\end{equation*}%
The above reminds one of formulas from \cite{AB_AP_JPA}.

This enables us to introduce the new system of generators $\{P_{\tau }, P_i, M_{\tau i}, M_{ij}\}\rightarrow\{\tilde P_{\tau }, \tilde P_i, M_{\tau i}, M_{ij}\}$ with
\begin{equation}\label{mr}
\tilde{P}_{\tau }\doteq \kappa \ln \Pi _{\tau },\qquad \tilde{P}_{i}\doteq
P_{i}\Pi _{\tau }^{-1}\quad \Rightarrow \quad \Pi _{\tau }=e^{\frac{\tilde{P}%
_{\tau }}{\kappa }}
\end{equation}%
which provides the deformed coproducts in the familiar Majid-Ruegg form
\begin{eqnarray}
\Delta _{\kappa }\left( \tilde{P}_{\tau }\right)  &=&1\otimes \tilde{P}%
_{\tau }+\tilde{P}_{\tau }\otimes 1,\qquad
\Delta _{\kappa }\left( M_{ij}\right) = 1\otimes M_{ij}+M_{ij}\otimes 1\label{MR_prym}\\
\Delta _{\kappa }\left( \tilde{P}%
_{i}\right)  &=&\exp(-\frac{\tilde{P}_{\tau }}{\kappa })\otimes \tilde{P}_{i}+%
\tilde{P}_{i}\otimes 1  \label{MR_Pi} \\
\Delta _{\kappa }\left( M_{\tau j}\right)  &=&M_{\tau j}\otimes 1+\exp(-\frac{
\tilde{P}_{\tau }}{\kappa })\otimes M_{\tau j}-\frac{1}{\kappa }\tau ^{2}%
\tilde{P}^{k}\otimes M_{kj}  \label{MR_N}
\end{eqnarray}
The algebraic relations in the Majid-Ruegg basis are
\begin{eqnarray}
\lbrack M_{\tau i},\tilde{P}_{\tau }] &=&-i\tau ^{2}\tilde{P}_{i} \quad,\quad   \lbrack M_{ij},\tilde{P}_{k}]=i(g_{jk}\tilde{P}_{i}-g_{ik}\tilde{P}_{j})  \quad,\quad\lbrack M_{ij},\tilde{P}_{\tau }]=0 \label{MRtau1} \\
\lbrack M_{\tau i},\tilde{P}_{j}] &=&\frac{i}{2}\kappa g_{ij}\left( 1-\exp(-\frac{2\tilde{P}_\tau}{\kappa})-\frac{\tau^2}{\kappa^2}\tilde{P}_i\tilde{P}^i\right)+\frac{i\tau ^{2}}{\kappa }\tilde{P}_{j}\tilde{P}_{i} \label{MRtau2}\end{eqnarray}
Notice that the Lie algebra of the stability group $G_\tau$ consist of the elements $\{M_{ij}\}$ for which the coproduct remains primitive.
The expressions (\ref{MR_prym})-(\ref{MRtau2}) cover all the standard $\kappa-$deformations; for the Lorentzian signature they describe both the time-like and the space-like quantizations.

\section{The null-plane (light-cone) deformation and the 2+(D-2) decomposition}

In the case of light-like deformation, i.e. when $\tau ^{2}=0$, one deals with the non-Euclidean geometry $IS\!O(p, q)$; $p, q\neq 0$.
Therefore we shall introduce the most convenient "light-cone" Poincar\'{e}
generators:
\begin{equation}\label{lc1}
P_{\mu }=\left( P_{+}\ ,P_{-}\ ,P_{a}\right) \quad, \qquad M_{\mu \nu }=\left(
M_{+\,-}\ ,M_{+\,a}\ ,M_{-\,a}\ ,M_{ab}\right)\quad, \qquad  a, b=1,2\ldots D-2.
\end{equation}%
as a basis in the Lie algebra $iso(g_{p, q})$. To this aim we have to
decompose the space $V^{D}=V^{2}\times V^{D-2}$, by a suitable choice of basic vectors,
into the orthogonal product of the two-dimensional Lorentzian space $\{V^{2},g_{AB}\}$ with a
 $D-2$ dimensional one $\{V^{D-2},g_{ab}\}$: $\left( A,B=+,-\right) $, $%
\left( a,b=1,2\ldots D-2\right) $. Moreover, the total metric $g_{\mu \nu }=g_{AB}\times g_{ab}$
becomes a product metric. We choose $g_{AB}=\left(
\begin{array}{cc}
0 & 1 \\
1 & 0%
\end{array}%
\right) $ in its anti-diagonal (light-cone) form as well as two null-vectors
$\tau ^{\mu }\equiv \tau _{+}^{\mu }=(1,0,\ldots 0)$, $\tilde\tau^\mu\equiv\tau _{-}^{\mu
}=(0,1,0\ldots 0)$: $\tau _{+}\tau _{-}=1$ in order to obtain the convenient
light-cone basis in the space of the Lie algebra generators (\ref{lc1}). This algebra
consists of the following (non-vanishing) commutators:
\begin{eqnarray}
\left[ M_{+\,a}\,,M_{-\,b}\right]  &=&-i\left( M_{ab}+g_{ab}M_{+\,-}\right)
\quad, \qquad  \left[ M_{\pm \,a}\,,M_{\pm \,b}\right] =0  \label{M+aM-b} \\
\left[ M_{\pm \,a}\,,M_{b\,c}\right]  &=&i\left( g_{ab}M_{\pm
\,c}-g_{a\,c}M_{\pm \,b}\right) \quad, \qquad  \left[ M_{+\,-}\,,M_{\pm \,a}\right]
=\pm iM_{\pm \,a}\   \label{MpmMbc} \\
\left[ M_{+\,-},P_{\pm }\right]  &=&\pm i\,P_{\pm }\quad, \qquad  \ \ %
\left[ M_{\pm \,a}\,,P_{b}\right] =ig_{ab}P_{\pm }\   \label{M_pmP} \\
\left[ M_{\pm \,a}\,,P_{\pm }\right]  &=&\left[ M_{+\,-}\,,P_{a}\right]
=0\quad, \qquad  \left[ M_{\pm \,a}\,,P_{\mp }\right] =-\,iP_{a}
\label{M_pmP_pm}
\end{eqnarray}%
together with the standard commutation relations within the $D-2$
dimensional sector $(M_{a\,b}\,,P_{a}\,,g_{ab})$, cf. (\ref{MM})--(\ref{PP}%
).
The universal formula for the coalgebra structure, in this case,  reduces to
\begin{eqnarray}
\Delta _{\tau }\left( M\right)  &=&M\otimes 1+1\otimes M\quad \qquad %
\mbox{for}\qquad M\in \{M_{+\,a}\,,M_{ab}\} \label{DtauM1}\\
\Delta _{\tau }\left( P\right)  &=&P\otimes \Pi _{+}+1\otimes P\quad \qquad %
\mbox{for}\qquad P\in \{P_{+}\,,P_{a}\}  \notag \\
\Delta _{\tau }\left( P_{-}\right)  &=&P_{-}\otimes \Pi _{+}+\Pi
_{+}^{-1}\otimes P_{-}-\frac{1}{\kappa }\left( P_{-}+\frac{1}{2\kappa }%
C_{+}\right) \Pi _{+}^{-1}\otimes P_{+}-\frac{1}{\kappa }P^{a}\Pi
_{+}^{-1}\otimes P_{a}\  \\
\Delta _{\tau }\left( M_{+\,-}\right) \, &=&M_{+\,-}\otimes 1+\Pi
_{+}^{-1}\otimes M_{+\,-}-\frac{1}{\kappa }P^{a}\,\Pi _{+}^{-1}\otimes
M_{+\,a} \\
\Delta _{\tau }\left( M_{-\,a}\right)  &=&M_{-\,a}\otimes 1+\Pi
_{+}^{-1}\otimes M_{-\,a}-\frac{1}{\kappa }\left( P_{-}+\frac{1}{2\kappa }%
C_{+ }\right) \Pi _{+}^{-1}\otimes M_{+\,a}-\frac{1}{\kappa }P^{b}\Pi
_{+}^{-1}\otimes M_{ba}  \label{T_M_-a} \label{DtauM2}
\end{eqnarray}%
where $\Pi _{+}\doteq 1+\frac{1}{\kappa }P_{+}$ and $\left( 1-\frac{1}{%
\kappa }P_{+}\Pi _{+}^{-1}\right) =\left( \Pi _{+}-\frac{1}{\kappa }%
P_{+}\right) \Pi _{+}^{-1}=\Pi _{+}^{-1}$ and $C_{+}$ is still to be
determined.  The Lie subalgebra corresponding to the stability group of $\tau _{+}$
consists of $iso(p-1, q-1)=gen\{M_{a\,b},M_{+\,b}\}$, i.e. the generators with the primitive coproducts.

On the other hand, the classical $r-$matrix corresponding to the vector $%
\tau _{+}$ reads
\begin{equation*}
r_{LC}=M_{+\,-}\wedge P_{+}+M_{+\,a}\wedge P^{a}
\end{equation*}%
Since $\tau _{+}^{2}=0$ it satisfies the CYB equation and generates the non-standard
(triangular) deformation. Its construction involves two Abelian $D-1$
dimensional subalgebras $\Gamma _{+}=gen\{M_{+\,-}\ ,P^{a}\}$ and $\Gamma
_{-}=gen\{P_{+}\ ,M_{+\,a}\}$ satisfying certain cross-commutation relations
(cf. formulas (\ref{M_pmP}-\ref{M_pmP_pm}) and Ref. \cite{Mudrov}). The
corresponding twisting element has the following form \footnote{%
It is called an extended Jordanian twist since it enlarges the basic Jordanian
twist $\exp \left( -iM_{+\,-}\otimes \ln \Pi _{+}\right) $ (see \cite{KLM}
for details).}:
\begin{eqnarray}\label{tw1}
\mathcal{F} &=&\exp \left( -iM_{+\,-}\otimes \ln \Pi _{+}\right) \,\exp \left( -%
\frac{i}{\kappa }M_{+\,a}\otimes P^{a}\Pi _{+}^{-1}\right)   \notag \\
&=&\exp \left( -\frac{i}{\kappa }M_{+\,a}\otimes P^{a}\right) \,\exp \left(
-iM_{+\,-}\otimes \ln \Pi _{+}\right)
\end{eqnarray}%
We are now in position to calculate coproducts directly from the twist by making use of the similarity transformation
\begin{equation}\label{tw_cop}
\Delta _{LC}(X)=\mathcal{F}\Delta _{0}(X)\mathcal{F}^{-1}
\end{equation}
 where $\Delta
_{0}(X)=X\otimes 1+1\otimes X$ denotes as before the primitive (undeformed) coproducts.
After performing the involved calculations it turns out that
\begin{equation}
  \mathcal{R}\Delta _{LC}\left( X\right)\mathcal{R}^{-1}=\Delta^{op} _{LC}\left( X\right)=\Delta _{\tau}\left( X\right)
\end{equation}
where $\mathcal{R}=\mathcal{F}_{21}\mathcal{F}^{-1}$ is a triangular quantum $R-$matrix, provided $C_+ = C$ as suggested by the formula (\ref{inv}). (Note that in the light-cone basis one has $C=2P_+P_- + P^aP_a$ and
$P_{-}+\frac{1}{2\kappa }C=P_{-}\Pi _{+}+\frac{1}{2\kappa }P^{a}P_{a}$). In other words, the formulas (\ref{DtauM1}) - (\ref{DtauM2}) and (\ref{tw_cop}) describe in a different way the same Hopf algebraic structure.

Another observation is that for coproducts $\Delta _{LC}\left(X\right)$ one can
introduce a partial analog of the Majid-Ruegg basis (observed before in \cite{0307038}). Indeed, setting $\tilde{P}%
_{+}=\ln \Pi _{+}\,,\tilde{P}_{a}=P_{a}\Pi _{+}^{-1}$ one has primitive coproduct  for  $\tilde{P}%
_{+}$ and  $M\in \{M_{+\,a}\,,M_{ab}\}$ as in (\ref{MR_prym}) and for $\tilde{P}_{a}$ and $M_{+-}$ as in (\ref{MR_Pi}) and (\ref{MR_N}) respectively. As far as algebra is concerned we get
\begin{eqnarray}
\left[ M_{+\,-},\tilde{P}_{+}\right] &=&i\left( 1-\exp(-\frac{\tilde{P}_{+}}{%
\kappa })\right) \quad, \qquad  \left[ M_{+\,a}\,,\tilde{P}_{b}\right]=
ig_{ab}\kappa \left( 1-\exp(-\frac{\tilde{P}_{+}}{\kappa })\right)  \\
\left[ M_{+\,a},\tilde{P}_{+}\right] &=&0\quad, \qquad  \left[ M_{+\,-}\,,\tilde{P}%
_{a}\right] =i\tilde{P}_{a}\left( 1-\exp(-\frac{\tilde{P}_{+}}{\kappa }%
)\right) \qquad, \qquad \left[ M_{-\,a}\,,\tilde{P}_{+}\right] =-\,\frac{i}{\kappa }%
\tilde{P}_{a}
\end{eqnarray}%
with the rest of commutators staying classical as in (\ref{M+aM-b})-(\ref{MpmMbc}). The only generator which does not fit
into this Majid-Ruegg scheme is $P_-\,$.\newpage

\section{Conclusions and perspectives}
In the study of Poincar\'{e} algebra, from the point of view of physical applications, one focuses on the representation theory and the value of Casimir operator $C=P^{2}$. We believe that physical objects are represented by time- or light-like four-momentum.
In this paper we have considered a coordinate analog of such four-vector in the $\kappa$-deformed case and we have shown that it is possible to consider analogous (time-, space- and light-like) cases which in fact parameterize the deformation. The universal formulas for the deformed Poincar\'{e} algebra depend on the choice of the additional vector field $\tau $ and allow one to consider three cases of deformations all being the symmetry of the corresponding noncommutative $\kappa(\tau)$-Minkowski spacetimes (\ref{kM}).
These non-equivalent deformations are classified by the stability groups of $\tau$. In other words we have presented a class of $\kappa-$deformations of orthogonal groups $S\!O(g)$ in the way that they explicitly depend on the choice of normalized four-vector $\tau$.

In the view of physical applications we can focus here on four dimensions with the metric of Lorentzian signature.  Deformed Hopf algebra describes a symmetry of  quantized  spacetime. In the most studied case the vector $\tau$ is time-like which corresponds to $\kappa-$Minkowski spacetime algebra with noncommutativity between time and space coordinates. Therefore such (time-like) vector can be identified with a preferred direction which can be interpreted as a four-velocity of the preferred observer. Corresponding $3+1$ decomposition provides a preferred frame. We have shown that in such frame the utilization of Majid-Ruegg coproducts is fully justified. It is also known that theories with preferred spacetime direction violate Lorentz invariance. In fact, the Lorentzian symmetry should be reduced to the stability subgroup for which the coproducts remain undeformed.
  
Also general relativity models with a preferred direction (Einstein-{\ae}ther) are currently under debate (see e.g. \cite{TJackobson} and references therein).  Astrophysical data indicates as well that the universe has a preferred (primordial) direction imprinted on the microwave background \cite{Ackerman}, \cite{Antoniou}. It has been already demonstrated that noncommutative effects turn out to be helpful for the explanation of this fact \cite{Balach1}, \cite{Balach2}, \cite{Naut}.

Alternatively, one can want to restore full Lorentz covariance  allowing to transform the vector $\tau$.
Indeed, the Lorentz transformation : $\tau^\mu\rightarrow\tilde\tau^\mu=\Lambda^\mu_\nu\tau^\nu$ does not change the orbit type. Therefore it preserves the deformation type. 
Such scenario is similar to that one which encounters in Special Relativity which admits a class of preferred observers (frames) - the inertial ones. Conversely, any two inertial observers are connected by the Lorentz transformation.  The same reasoning allows one to restore full diffeomorphism invariance on a curved background.

Another nice feature of our approach comes from the fact that the  metric tensor $g$ determining the orthogonal group does not need to be in its canonical (diagonal) form. It implies that the deformation can be executed in an arbitrary coordinate system. In particular, on a curved manifold when the components of the metric $g$ representing gravitational field as well as the components of $\tau$ are point-wise dependent. In such a case the usage of Majid-Ruegg bases is not, in general, allowed unless some stronger assumptions (e.g. global hyperbolicity, foliation, etc.) are taken into account. The unified description seems to be particularly useful from the point of view of applications in
deformations of general relativity. A deformation in the geometric setting has been under investigation for quite some time as an alternative to the quantization of gravity \cite{Wess}. For example one can follow the most recent proposition to consider gravitational and cosmological models induced directly from noncommutative, i.e. quantum spacetime \cite{Majid-Beggs}. In this approach the metric is taken to be a function of coordinates. The
unified description already suitable for the arbitrary metric tensor could be generalized to include the metric as a function of coordinates belonging to the center of algebra (\ref{kM}). Such generalization would require suitable modification of the
Poincar\'{e} algebra (\ref{PP}) (see e.g. \cite{DK-SM-AP-RS_PLB}). However, the quantum deformation would be still described by the classical r-matrix $r_\tau$ and would imply two possibilities for the deformation, i.e. $\tau^2\neq 0$ or $\tau^2=0$. This could allow to obtain more gravitational and cosmological models induced by (\ref{kM}).
Another application could be found in the so-called relative-locality where the metric might live on the momentum space \cite{GAC}. 
The so-called relative locality effects were already investigated in the
time-like case \cite{GAC1,GAC2}. The unified description, however, opens a way to also
consider the light-like one. In a similar fashion as used in \cite{1210-6814}
one could define momenta realizations compatible with light-like
deformation, i.e. twisted realizations and proceed with the relative-locality formulation \cite{GAC1}.
Moreover, the unified description proposed in this paper might be of use in the formulation of quantum field theory on Lie algebraic type of noncommutative spacetimes; see e.g. \cite{SM_JHEP}, \cite{LukWor}.

\section*{Acknowledgements}

This work is a part of the Polish National Science Centre (NCN) project 2011/01/B/ST2/03354.
AB acknowledges the financial support from Erasmus Mobility and
Training Program under the contract no. FSS/2013/MOB/W/0038 as well as the Science Institute,
University of Iceland for the hospitality. The authors thank Jerzy Lukierski and Stjepan Meljanac for remarks and discussions.

\end{document}